\documentclass[aps,prl,twocolumn,groupedaddress,showpacs]{revtex4}
\usepackage{graphicx}
\usepackage{color}
\usepackage{amsmath}
\usepackage{amssymb}

\begin{document}
\title{Boson Mott insulators at finite temperatures}
\author{Fabrice Gerbier}
\affiliation{Laboratoire Kastler Brossel, ENS, Universit\'e Pierre
et Marie-Curie-Paris 6, CNRS ; 24 rue Lhomond, 75005 Paris, France}
\email{fabrice.gerbier@lkb.ens.fr}
\date{\today}
\begin{abstract}
We discuss the finite temperature properties of ultracold bosons in
optical lattices in the presence of an additional, smoothly varying
potential, as in current experiments. Three regimes emerge in the
phase diagram: a low-temperature Mott regime similar to the
zero-temperature quantum phase, an intermediate regime where MI
features persist, but where superfluidity is absent, and a thermal
regime where features of the Mott insulator state have disappeared.
We obtain the thermodynamic functions of the Mott phase in the
latter cases. The results are used to estimate the temperatures
achieved by adiabatic loading in current experiments. We point out
the crucial role of the trapping potential in determining the final
temperature, and suggest a scheme for further cooling by adiabatic
decompression.
\end{abstract}
 \maketitle
%
%
%%%%%%%%%%%%%%%%%%%%%%%%%%%%%%%% body %%%%%%%%%%%%%%%%%%%%%%%%%%%%%%
%
%
Ultracold gases in optical lattices are currently a major topic in
the field of ultracold atoms, both experimentally and theoretically
(see the reviews
\cite{jaksch2005a,morsch2006a,lewenstein2006a,bloch2007a}. So far,
experiments are mostly carried out with Bose gases, and interpreted
in terms of the zero-temperature phase diagram of the Bose-Hubbard
model \cite{jaksch2005a,lewenstein2006a,bloch2007a}. However, actual
experiments inevitably take place at finite temperatures, which
leads, {\it e.g.}, to residual number fluctuations. In view of its
importance for applications such as the controlled generation of
entangled states \cite{jaksch2005a}, or the study of quantum
magnetism \cite{lewenstein2006a}, this issue has recently received
increasing interest
\cite{dickerscheid2003a,plimak2004a,demarco2005a,rey2006a,pupillo2006a,schmidt2006a,capogrosso2007a,lu2006b,ho2007a}.
In this paper, we first discuss how the phase diagram of a ultracold
Bose gas in an optical lattice is modified at finite temperatures.
We identify a melting temperature $T^\ast\approx 0.2~U/k_{\rm B}$,
above which the system is entirely thermal, and a much lower
critical temperature $T_{\rm c}\approx zJ$, above Mott regions
survive, but superfluidity is absent. Here $z$ is the coordination
number, $U$ the on-site interaction energy, and $J$ the tunneling
energy. We derive analytically the thermodynamics of the Mott phase
for $T<T^\ast$, including particle-hole corrections. Finally, we
model the procedure adopted in current experiments to produce
quantum gases in optical lattices, where a Bose-Einstein condensate
is slowly loaded into the optical potential in the presence of an
``external'', smoothly varying trap. Assuming the loading is done
adiabatically, we estimate the final temperatures that can be
reached with this procedure. We point out the crucial role of the
external potential in determining the final temperature, and suggest
a scheme for further cooling by adiabatic decompression.

\begin{figure}
%\resizebox{\columnwidth}{!}{
\includegraphics[width=8cm]{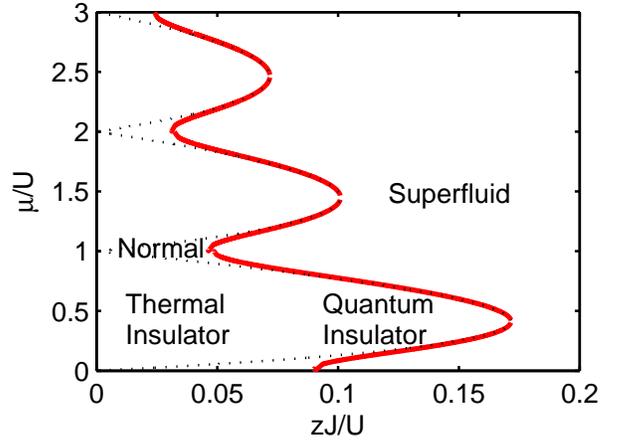}
%}
\caption{Phase diagram for an ultracold Bose gas in an optical
lattice at finite temperatures ($k_B T=U/20$). The finite-
temperature Mott lobes boundaries are shown as solid lines, and the
zero-temperature boundaries as dotted lines for comparison. With
increasing temperature, the superfluid regions in between the Mott
lobes shrink in width and shift to larger tunneling energies. For
temperatures $zJ<T$, the SF-MI transition is replaced by a smooth
crossover between two Mott domains across a normal phase. This
generic phase diagram persists up to the melting temperature
$T^\ast\approx0.2~U$ where the Mott domains vanish.}
\label{phasediagram}
\end{figure}

For future reference, we first recall the phase diagram for zero
temperature. For $J=0$, the ground state is a product of identical
Fock states, with $n_0$ atoms at each site depending on the chemical
chemical potential $\mu$ according to $n_0={\rm Integer}(\mu/U)+1$.
When a weak tunneling between nearest neighbors is allowed, the
ground state remains a Mott insulator with suppressed number
fluctuations but only in well-defined domains (``Mott lobes'') in
the $J-\mu$ plane (dashed lines in Fig.~\ref{phasediagram}). Outside
these domains, the MI is unstable towards a delocalized, superfluid
state (SF), that connects to a condensate in the Bloch state ${\bf
q=0}$ when $J \gg U$. In current experiments, an external harmonic
potential of the form $V_{\rm T}({\bf r})=\frac{1}{2}m\omega_{\rm
T}^2r^2$ is usually superimposed on the lattice potential. As a
result, the density profile is in general not uniform everywhere,
but has the shape of a ``wedding cake'' formed by successive layers
(``Mott shells'') with integer densities \cite{foelling2006a}. If
this potential varies slowly over a few lattice sites, the local
density approximation (LDA) can be employed. It obtains the
(coarse-grained) number $n_{\rm h}[\mu,T]$ and entropy $s_{\rm
h}[\mu,T]$ densities calculated as in the homogeneous case, but for
a local chemical potential $\mu_0-V_{\rm T}({\bf r})=\mu_{\rm
loc}({\bf r})$. In the following, we first compute the phase diagram
for uniform systems, and use the LDA to describe trapped gases.

Within a Mott domain with $n_0$ bosons per site, the lowest-lying
excited states are ``particle'' and ``hole'' states, where a
supplementary particle is added (respectively removed) to the
background value $n_0$ with a free energy cost $U n_0-\mu$ (resp.
$\sim\mu-U(n_0-1)$). Other occupation numbers correspond to free
energies at least of order $U$, and are therefore suppressed by a
thermal factor $\sim e^{-\beta U}\ll 1$. This allows to use what we
call the ``particle-hole approximation'' (PHA)
\cite{rokshar1991a,krauth1992a,altman2001a,gerbier2005b,pupillo2006a,rey2006a},
which consists in truncating the on-site Hilbert space to the states
$n_0$, $n_0-1$ and $n_0+1$. As we will see, this approximation,
valid for temperatures $k_B T \ll U$ , allows to describe quite
extensively the physics of the insulating phase at finite
temperatures.

\begin{figure}
%\resizebox{\columnwidth}{!}{
\includegraphics[width=8cm]{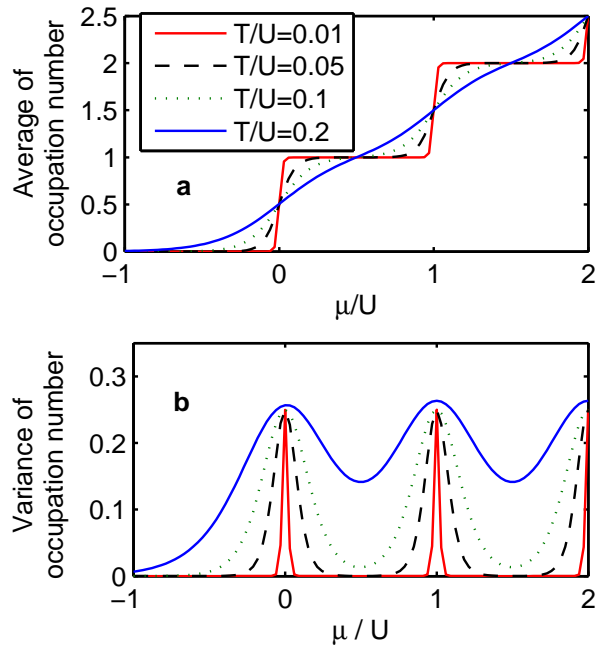}
%}
\caption{Density profile and fluctuations for an array of isolated
wells ($J=0$), shown for various temperatures. }
\label{densityprofile}
\end{figure}

As a starting point, we first neglect tunneling between the wells
completely and consider the thermodynamics of a uniform system of
isolated wells, where the atoms interact locally through the on-site
interaction. Since the wells are independent, the global partition
function factorizes into a product of identical on-site partition
functions $z_0=\sum_n \exp\left[-\beta(E(n)-\mu n)\right]$, where
$E(n)=U n(n-1)/2$ and $\beta=1/k_{\rm B}T$. Keeping only the
particle/hole excitations with energies $E_0^{\rm (qp)}=U n_0$ and
$E_0^{\rm (qh)}=U (n_0-1)$, the mean density and variance of density
fluctuations take the simple forms
\begin{eqnarray}
\overline{n}_0 & \approx n_0 + \left(B^{(+)}-B^{(-)}\right)/z_0,\\
\mathrm{Var(n)}_0 &\approx \left(B^{(+)}+B^{(-)}\right)/z_0^2,
\end{eqnarray}
where $B^{(+)}=e^{\beta[\mu-U n_0]}$ and $ B^{(-)}=e^{\beta[U
(n_0-1) - \mu]}$ are Boltzmann factors corresponding respectively to
adding or removing a particle from the ``background'' value $n_0$,
and where $z_0 = 1+B^{(+)}+B^{(-)}$. In Fig.~\ref{densityprofile},
we have shown the density and fluctuations as a function of the
chemical potential. With increasing temperatures, the step-like
density profile characteristic of a Mott state becomes increasingly
smoother, reflecting the fact that the free energy cost to create
supplementary particles or holes vanishes near the edges of a Mott
plateau. A Mott-like region survives up to $T^\ast \sim 0.2~U/k_B$
(see also \cite{demarco2005a}), which can be seen as a melting
temperature for the Mott phase. We have also compared the exact
expressions to the PHA, without noticeable difference up to
$T^\ast$.

With the properties of the zero-tunneling model clarified, we
reintroduce tunneling in a second step. At zero temperature, when
exploring a Mott domain at constant $J$ and varying chemical
potential, the Mott state remains stable as long as its elementary
(quasi-particle and hole) excitations are. At the upper or lower
Mott lobe boundaries, the chemical potential $\mu^{(\pm)}$becomes
equal to the excitations energies $E_{{\bf k}=0}^{\rm (\rm qp/qh)}$,
thus favoring the proliferation of particle or hole excitations to
reduce the free energy \cite{elstner1999a}. To investigate how this
behavior changes with increasing temperature, we have calculated the
quasiparticles and holes dispersion relations at finite temperature
using the PHA supplemented by a random-phase approximation as in
\cite{vanoosten2001a,dickerscheid2003a,gangardt2006a,sengupta2005a}.
We find
%the $J=0$ Matsubara Green function is given in the PHA by
%\begin{eqnarray}
%G_0(i \omega_n) \approx \frac{(n_0+1)A^{({\rm qp})}}{i \hbar\omega_n
%+\mu - U n_0} - \frac{n_0 A^{({\rm qh})}}{i \hbar\omega_n +\mu - U
%(n_0-1)}.
%\end{eqnarray}
%Finite temperature effects are contained in $A^{(\rm qp/qh)}=(1 -
%B^{(\pm)})/z_0$. For a finite tunneling, the Green function can be
%calculated from $G_0$ using a random phase approximation
%\cite{vanoosten2001a,dickerscheid2003a,gangardt2006a,sengupta2005a}.
%After some algebra, one can obtain the dispersion relations
%explicitly.
%\begin{equation}\label{green}
%\frac{1}{\hbar}G({\bf k},i \omega_n)=\frac{Z_{\bf k}}{i\hbar
%\omega_n +\mu -E_{\bf k}^{(\rm qp)}}+\frac{1-Z_{\bf k}}{i\hbar
%\omega_n + \mu-E_{\bf k}^{(\rm qh)}},
%\end{equation}
%but with modified dispersion relations,
\begin{eqnarray}
E_{\bf k}^{(\rm qp/qh)} & = & \frac{C J_{\bf
k}}{2}+U(n_0-\frac{1}{2})\pm \Delta_{\bf k},\label{EkpmT}
\end{eqnarray}
with $\Delta_{\bf k}=\sqrt{C^2 J_{\bf k}^2+ 2 U D J_{\bf k}+U^2}$,
$J_{\bf k}=-2 z J \chi_{\bf k}$ and $\chi_{\bf
k}=\frac{1}{z}\sum_{i=1}\cos(k_i d)$. The finite temperature
dispersion relations are formally similar to the zero temperature
expressions \cite{vanoosten2001a} except for the factors $C/D =
(n_0+1) A^{({\rm qp})} \mp n_0 A^{({\rm qh})}$, with $A^{(\rm
qp/qh)}=(1 - B^{(\pm)})/z_0$.

In the following, we mostly work for simplicity in the limit $U\gg
zJ$, where the dispersion relations reduce to $E_{\bf k}^{(\rm
qp/qh)} \approx E_{0}^{(\rm qp/qh)} \mp J^{(\rm qp/qh)} \chi_{\bf
k}$, with an effective tunneling energy $J^{(\rm qp)} =z J (n_0+1)
A^{(\rm qp)}$ for particles and $J^{(\rm qh)} =z J n_0 A^{(\rm qh)}$
for holes. We concentrate first on the upper boundary and try to
find the chemical potential for which $E_{{\bf k}=0}^{(\rm
qp)}=\mu^{(+)}$. Introducing a variable $s=(U n_0 - \mu^{(+)})/k_B
T$, we find that, up to terms $\sim e^{-\beta U}$, $s$ solves the
equation $s - a \tanh\left( \frac{s}{2}\right)=0$, with $a=z J
(n_0+1)/k_B T$. Graphic inspection shows that there is no solution
for $a<2$, corresponding to a critical temperature $k_B T_c= z J
(n_0+1)/2$ above which there is no SF region left due to thermal
depletion. This argument holds far from the zero-temperature Mott
transition, where Eqs.~(\ref{EkpmT}) are valid. To obtain the
complete phase boundaries, we have solved numerically the equations
$E_{{\bf k}=0}^{(\rm qp/qh)}=\mu^{(\pm)}$ for any $z J/U$. The
result is shown in Fig.~~\ref{phasediagram}, where three distinct
regions emerge. For $zJ  \ll 2 k_B T/(n_0+1)$, one goes continuously
from one insulating ``lobe'' to the other through a region in the
normal phase. For $ 2 k_B T/(n_0+1) \ll zJ \ll z J_c$, well-defined
Mott lobes survive, as in the ground state, and above the critical
$J_c$ a phase transition takes place to a SF phase, albeit with
reduced SF fraction. Above $T\sim T^\ast$, we expect the system to
be essentially normal for $J<U$. Note that when Mott regions with
different filling factors due to an external trap potential, the
regimes described above can coexist.

To characterize the thermodynamics for $T < T^\ast$ more
quantitatively, we calculate the free energy as
%\begin{equation}\label{F1}
%\frac{F}{k_B T}=\sum_{{\bf k},i\omega_n} \ln \left( \frac{G_0 (i
%\omega_n)}{G(i \omega_n, {\bf k})}\right).
%\end{equation}
%The sum over Matsubara frequencies can be performed explicitly by
%contour integration and gives
$F=F_0+\Delta F^{(\rm qp)}+\Delta F^{(\rm qh)}$. The first term
$F_0=-k_B T \ln(1+B^{(+)}+B^{(-)})$ gives the free energy
corresponding to the thermal activation of local defects in each
well for zero tunneling. The remaining contributions $\Delta F^{(\rm
qp)}  =F_{\rm BE}\left[E_{\bf k}^{(\rm qp)}-\mu\right]-F_{\rm
BE}\left[U n_0-\mu\right]$ and $\Delta F^{(\rm qh)}  =F_{\rm
BE}\left[\mu-E_{\bf k}^{(\rm qh)}\right]-F_{\rm
BE}\left[\mu-U(n_0-1)\right]$ describe instead dilute gases of
quasi-particles and quasi-holes mobile through the lattice. Here,
the functional $F_{\rm BE}[x]=\sum_{{\bf k}} \ln \left(1- \exp(\beta
x)\right)$ gives the free energy of an ideal Bose gas. %The $F_{\rm BE}\left(U n_0\right)$
%and $F_{\rm BE}\left(U (n_0-1)\right)$ ensure that the free energy
%reduces to $F_0$ ias $J\rightarrow 0$.
For $U \gg J$, we obtain for instance
\begin{eqnarray}\label{deltaFqplowJ}
\frac{\Delta F^{( \rm qp)}}{N_s k_B T}& = & -\sum_{m=0}^\infty
\frac{e^{m \beta (\mu-U n_0)}}{m}\left[ I_0 \left(\frac{m \beta
J^{(\rm qp)}}{\mathcal{D}}\right)^\mathcal{D}-1\right],
\end{eqnarray}
and a similar expression for $\Delta F^{( \rm qh)}$. Here $I_0$ is a
Bessel function, and $\mathcal{D}$ the dimensionality.

Thermodynamic quantities follow from derivatives of the free energy.
Contributions from local activation, denoted by the index "$ _0$"
have been calculated above. Terms coming from particle and hole
contributions are readily obtained. For instance, the particle/hole
corrections to the average density read $\Delta \overline{n}^{(\rm
qp/qh)}=-
\partial \Delta F^{( \rm qp/qh)}/\partial \mu$, whereas the density fluctuations can be characterized using the
Hellmann-Feynman theorem \cite{kheruntsyan2003a}, which to leading
order in $e^{-\beta U}$, reduces to ${\rm Var}(n) \approx {\rm
Var}(n)_0+ \Delta \overline{n}^{(\rm qp)}  ( 1-\Delta
\overline{n}^{(\rm qp)}) - \Delta \overline{n}^{(\rm qh)} ( 1+\Delta
\overline{n}^{(\rm qh)})$.

In the low-temperature limit, $T \ll T_c$, the contribution of local
activation terms is negligible, and tunneling corrections to the
thermodynamic functions dominate. However, using the asymptotic form
for $I_0$ we find for instance $\Delta F^{(\rm qp)}/N_s k_B T = [k_B
T/4\pi J (n_0+1)]^{3/2} g_{5/2} \left(
e^{\beta(\mu-\mu^{(+)})}\right)$, where $g_\alpha(x)=\sum_n x^n
/n^\alpha$ denotes a Bose function. The prefactor implies that
corrections to the zero-temperature behavior are highly suppressed.
Hence this regime correspond to a quantum Mott insulator, where the
properties of the zero-temperature system are essentially preserved
up to small thermal corrections. Increasing the temperature,
corrections from local activation and from tunneling terms become
progressively comparable until the high temperature regime, $T \gg
T_c$, is reached. In this regime, using the Taylor expansion
$I_0(x)\approx 1+x^2/4$ for the Bessel function, we find
\begin{eqnarray}\label{deltaFqphighT}
\frac{\Delta F^{(\rm qp)}}{N_s k_B T}& \approx & -\frac{1}{4
\mathcal{D}}\left( \frac{J(n_0+1)}{k_B T} \right)^{2}
\frac{B^{(+)}}{z_0^2}.
\end{eqnarray}
Note that this expression also applies to the normal phase
in-between two Mott-like regions. Here the right hand side has a
small term $\propto (J n_0/k_B T)^2 \ll 1$, times a factor
comparable to the $J=0$ result. This means that in this temperature
regime, the model with $J=0$ alone already gives a good description
of the thermodynamics.

\begin{figure}
%\resizebox{\columnwidth}{!}{
\includegraphics[width=8cm]{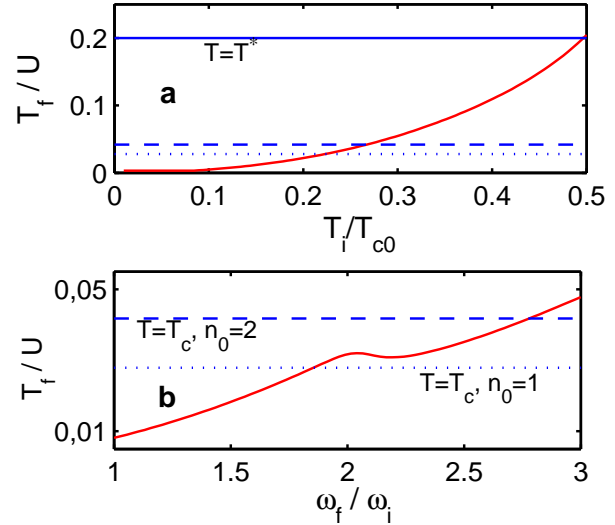}
%}
\caption{Final temperature achieved by adiabatically transferring a
condensate into the optical lattice potential (depth $V_0=20~E_{\rm
R}$). The transfer is done in {\bf (a):} for a fixed final trap
frequency $\omega_f=2\pi \times 70~$Hz varying the initial
temperature $T_i$, and in {\bf (b):} for a fixed initial temperature
$T_i=0.3~T_{\rm c0}$ but varying the final trap frequency
$\omega_f$. The solid line indicates the melting temperature
$T^\ast$. Above the dashed and dotted lines, the superfluid layers
corresponding to $n_0=2$ and $n_0=1$, respectively, turn normal. For
both plots, the initial frequency $\omega_i=2\pi\times20~$Hz and the
atom number $N=2\times 10^5$ are fixed.} \label{adiabatic_transfer}
\end{figure}

As explained above, the results can be directly applied to the
harmonically trapped case in the LDA. We apply now the calculation
to estimate the temperatures achieved in current experiments.
Typically, a Bose-Einstein condensate is first produced in a
magnetic trap with frequency $\omega_{ i}$, at an initial
temperature $T_i$. We calculate its entropy using the Popov approach
as exposed in \cite{Giorgini1996a} using the local density
approximation and for typical experimental values
$\omega_i=2\pi\times 20~$Hz and $N= 2 \times 10^5$ atoms. Then, the
cloud is slowly transferred in the optical lattice of depth $V_0$.
This also changes the frequency $\omega_f$ of the external harmonic
potential according to
\begin{equation}\label{omegaf}
\omega_f=\sqrt{\omega_{i}^2+\frac{8 V_0}{m w^2}}.
\end{equation}
In Eq.~(\ref{omegaf}) the second term comes from the trapping force
due to the Gaussian shape of the laser beams forming the lattice,
with $w$ the lattice laser waist ($1/e^2$ radius), typically
$\approx~150~\mu$m. Assuming that the transfer into the lattice is
adiabatic, both entropy and atom number are conserved. In the final
state, assumed to be deep in the MI regime, the number density
$n=-\partial f/\partial \mu$ and entropy density $s=-\partial
f/\partial T$ are calculated from the free energy density $f$
\footnote{For this calculation, tunneling corrections are calculated
using Eq.~(\ref{deltaFqphighT}) rather than the full
Eq.~(\ref{deltaFqplowJ}).}. Those densities are then integrated over
space to fix the total atom number and entropy, which determines the
final chemical potential and final temperature $T_f$. We plot in
Fig.~\ref{adiabatic_transfer}a the result of a self-consistent
calculation solving for $T_f$ for a fixed atom number
$N=2\times10^5$. One sees that for small initial temperature
$T_i<0.4~T_{\rm c0}$ (condensed fraction larger than 75 \%) one ends
well within the Mott region. Only by starting from a rather large
thermal fractions does the final temperature exceed $0.2~U$.
Reaching the quantum insulator regime is more challenging, but still
within reach of current experimental possibilities ($T_i<0.2~T_{\rm
c0}$).

For this calculation, the role of the trapping potential is
essential. Similar calculations for homogeneous systems with integer
filling would yield a much higher temperature \cite{schmidt2006a},
since near the center of the Mott lobe the entropy is exponentially
suppressed by the finite interaction energy $U$. In the trapped
case, the entropy in fact concentrates around the lobe boundaries,
where the generation of excitations is easiest. As long as $T <
T^\ast$, the peak value of the entropy density is  $s_{\rm
max}\approx k_{\rm B}\log(2)$. The thickness $\delta R$ of the layer
where thermal fluctuations are important can be estimated from $m
\omega_{\rm T}^2 R_0 \delta R \sim k_{\rm B} T$, where
$R_0=\sqrt{2U/m\omega_{\rm T}^2 d^2}$ is a typical spatial scale for
the shell structure. Since the total entropy and atom number scale
respectively as $S/ k_{\rm B}\sim 4\pi R_0^2 \delta R$ and $N\sim
(4\pi/3)(R_0/d)^{3/2}$, we find that in the trappe case, the entropy
per particle rises linearly with temperature, $S/N k_{\rm B}\sim 3
k_{\rm B} T/U$, instead of exponentially ($\sim e^{-U/k_{\rm B }T}$)
as in the uniform case (see also \cite{ho2007a} for a similar
analysis).

An important consequence of Eq.~(\ref{omegaf}) is that the trap
frequency increases significantly when the lattice depth is raised.
This leads to spatial compression (and the formation of Mott
plateaux with higher filling), but also to adiabatic heating of the
cloud, as in a conventional harmonic trap. In the second
calculation, this effect was investigated \footnote{This effect was
independently noticed and analyzed in \cite{ho2007a}.}. In
Fig.~\ref{adiabatic_transfer}b, the result is plotted for a fixed
initial temperature $T_i=0.3~T_{\rm c0}$, but varying final trap
frequencies. It is clear that lower and lower temperatures are
achieved when the external trap frequency is reduced. In particular,
one can achieve $T<0.01~U$ if the transfer is done at constant trap
frequency. Experimentally, this could be realized using blue-detuned
lattice beams, or additional blue-detuned (repulsive) lasers to
compensate for the trapping force of the laser beams. Note finally
the bump in the curve, that indicates that the transition from a
unity-filled MI to a two-plateaux distribution with either single or
double occupancy.

In conclusion, we have discussed the thermodynamics of ultracold
bosons in optical lattices. We have identified two characteristic
temperatures, $T \sim T_c$, above which the SF regions disappear,
and a melting temperature $T^\ast\approx 0.2~U$. The thermodynamic
function for the Mott phases are calculated explicitly. We have used
the results to estimate the temperature reached in current
experiments, and found that they easily reach the thermal insulator
regime $T<T^\ast$, and possibly the quantum region $T < T_c$. We
suggest an adiabatic decompression scheme which potentially allows
to reach much lower temperatures in a system with unity filling.

I would like to thank Immanuel Bloch, Jean Dalibard, Simon
F\"{o}lling, Dries van Oosten and Artur Widera for discussions and
comments. I acknowledge support from IFRAF and ANR.

%\bibliographystyle{apsrev}
%\bibliography{mott_2}

\end{document}